\documentclass[twocolumn,floatfix,superscriptaddress,a4paper,showkeys,nofootinbib,reprint]{revtex4}
\usepackage[colorlinks=true,linktocpage=true,linkcolor=blue,citecolor=blue,allcolors=blue]{hyperref}
\usepackage{epsfig}
\usepackage{latexsym}
\usepackage[utf8]{inputenc}
\usepackage{xspace}
\usepackage{indentfirst}
\usepackage{enumitem}
\usepackage{color}

\usepackage{setspace}% http://ctan.org/pkg/setspace
\usepackage{lipsum}% http://ctan.org/pkg/lipsum

\usepackage{amsmath}
\usepackage{amssymb}
\usepackage[english]{babel}
\usepackage{url}

\newcommand{\Eq}[1]{Eq.~(\ref{#1})}

\newcommand{\eq}[1]{\begin{align} #1 \end{align}}

\begin{document}

\title{
The analytic structure of thermodynamic systems with repulsive interactions
}

\author{Kirill Taradiy}
\affiliation{Bogolyubov Institute for Theoretical Physics, 03680 Kiev, Ukraine}
\author{Anton Motornenko}
\affiliation{Institut f\"ur Theoretische Physik,
Goethe Universit\"at Frankfurt, D-60438 Frankfurt am Main, Germany}
\affiliation{Frankfurt Institute for Advanced Studies, Giersch Science Center,
D-60438 Frankfurt am Main, Germany}
\author{Volodymyr Vovchenko}
\affiliation{Institut f\"ur Theoretische Physik,
Goethe Universit\"at Frankfurt, D-60438 Frankfurt am Main, Germany}
\affiliation{Frankfurt Institute for Advanced Studies, Giersch Science Center,
D-60438 Frankfurt am Main, Germany}
\author{Mark I. Gorenstein}
\affiliation{Bogolyubov Institute for Theoretical Physics, 03680 Kiev, Ukraine}
\affiliation{Frankfurt Institute for Advanced Studies, Giersch Science Center, D-60438 Frankfurt am Main, Germany}
\author{Horst Stoecker}
\affiliation{Institut f\"ur Theoretische Physik,
Goethe Universit\"at Frankfurt, D-60438 Frankfurt am Main, Germany}
\affiliation{Frankfurt Institute for Advanced Studies, Giersch Science Center,
D-60438 Frankfurt am Main, Germany}
\affiliation{GSI Helmholtzzentrum f\"ur Schwerionenforschung GmbH, D-64291 Darmstadt, Germany}

\date{\today}

\begin{abstract}
Thermodynamic properties of systems with repulsive interactions, 
are considered in the grand canonical ensemble. 
The analytic structure of the excluded-volume model
in the complex plane of the system chemical potential~(fugacity) is elaborated, based on the fact that the pressure function can be given in terms of the Lambert W-function.
Even though the excluded volume model has no phase transitions at real values of the chemical potential, it does exhibit a branch cut singularity in the complex plane, thus limiting the convergence range of the Taylor expansion in the chemical potential.
Close similarities to analytic properties of the other models with repulsive interactions, such as a cluster expansion model, the mean-field model, and the ideal Fermi gas model, are pointed out.
As an example, repulsive baryonic interactions in a hadron gas, with a focus on the fugacity/virial and Taylor expansion methods used in lattice QCD, are presented.
The asymptotic behavior of the Fourier expansion coefficients in these various models suggests that the singular part of net baryonic density can to leading order be universally expressed in terms of polylogarithms.
\end{abstract}
% \pacs{15.75.Ag, 24.10.Pq}


\maketitle

%\begin{spacing}{1.9}

\section{Introduction}

Properties of strongly interacting matter and determination of its different phases are the key questions which drive the heavy-ion collision experiments as well as finite-temperature lattice QCD simulations.
The first-principle lattice methods are restricted to simulations at zero or imaginary chemical potentials due to the sign problem.
The indirect lattice methods to probe finite baryon densities are based on extrapolations such as analytic continuation from imaginary
chemical potential~\cite{deForcrand:2002hgr,DElia:2002tig,Gunther:2016vcp} or the Taylor
expansion method~\cite{Allton:2002zi,Gavai:2008zr,Kaczmarek:2011zz,Bazavov:2017dus}.
Both methods are sensitive to the analytic properties of the pressure function in the complex chemical potential plane, in particular its singularities.
These restrict the scope of the analytic continuation as well as the convergence radius of Taylor expansion.
Knowledge of the possible singularities is thus useful to control the validity and accuracy of both these methods.
Often the singularities of the pressure function are associated with  phase transitions and critical phenomena.
Important examples include the chiral phase transition in the chiral limit of QCD~\cite{Pisarski:1983ms,Stephanov:2006dn} and a suspected critical point at finite baryon density~\cite{Stephanov:2006dn}.
As we show below, however, the pressure function singularities are not necessarily connected to physical phase transitions.

Useful guidance is provided by phenomenological models, which incorporate various symmetries and physical mechanisms expected in a given region of the phase diagram.
Here we employ hadron resonance gas (HRG) models which are used to provide a reasonable description of the hadronic part of the QCD-matter phase diagram. 
A wide range of HRG applications includes the description of hadron yields in heavy-ion collisions~[see, e.g.,~Refs.~\cite{Becattini:2009sc,Braun-Munzinger:2015hba} for a review] and lattice QCD data at moderate temperatures~\cite{Borsanyi:2013bia,Bazavov:2014pvz}.
Common extensions of the ideal HRG model include the incorporation of the repulsive  interactions~\cite{Olive:1980dy,Rischke:1991ke,Venugopalan:1992hy,Yen:1997rv}.
The relevance of repulsive baryonic interactions in the HRG equation of state has recently been established through an analysis of the lattice gauge theory data on baryon number susceptibilities at zero chemical potentials~\cite{Vovchenko:2016rkn,Huovinen:2017ogf} and on Fourier coefficients of net baryon density at imaginary baryonic chemical potential~\cite{Vovchenko:2017xad} as well baryon number fluctuations in heavy-ion collisions~\cite{Fukushima:2014lfa,Albright:2015uua}.

In this paper we study the analytic properties of these models in the complex chemical potential plane.
Certain related features, such as the distribution of the Lee-Yang zeros, have been studied within the excluded volume~(EV) and mean field~(MF) models long time ago~\cite{hauge1963yang}.
Here we present a more complete picture, using the fact that the grand canonical thermodynamic functions in these two models can be expressed in terms of the Lambert W-function.
We consider also the cluster expansion model from Ref.~\cite{Vovchenko:2017gkg} and the ideal Fermi gas, in addition to the EV and MF models. 
All these distinct models are found to exhibit a quite similar analytic structure of their grand-canonical thermodynamic potentials.
The results are discussed in light of possible applications to a reasonable analysis of lattice QCD data.

The paper is organized as follows.
Section~\ref{sec:EVsingle} goes in detail through the analytic solution of a single-component Maxwell-Boltzmann gas with
EV interactions.
Section~\ref{sec:evhrg} explores the analytic properties of the HRG with EV interactions.
Section~\ref{sec:othermodels} compares a number of distinct HRG models with repulsive interactions.
Summary in Sec.~\ref{sec:summary} closes the article.

\section{Single component excluded volume model}
\label{sec:EVsingle}

The pressure of a single-component Maxwell-Boltzmann gas with a van der Waals-type EV correction is given by
\eq{\label{ev}
p~=~\frac{n T}{1-bn}~,
}
where $T$ and $n$ are the system's temperature and particle number density, respectively, and $b$ is the excluded volume parameter.
The pressure (\ref{ev}) in the grand canonical ensemble (GCE) is presented
in terms of the following transcendental equation~\cite{Rischke:1991ke}:
\eq{\label{eq:evol}
p(T, \lambda) = T \, \phi(T) \, \lambda \, \exp\left(- \frac{b \,p}{T} \right).
}
Here
$\lambda = \exp(\mu/T)$ is the fugacity, $\mu$ is the chemical potential, and
\eq{
\phi(T) & = \frac{d \, m^2 \, T}{2 \pi^2} \, K_2(m/T).
}
Here $d$ and $m$ are particle's degeneracy factor and mass, respectively, and $K_2$ is the
modified Bessel function.
The pressure plays the role of the thermodynamical potential in the GCE, $T$ and $\mu$ are the corresponding independent intensive variables. All thermodynamical functions can be calculated
in terms of $p(T,\mu)$ and its partial derivatives.

The solution of \Eq{eq:evol} can be written explicitly as~\cite{NoronhaHostler:2012ug}:
\eq{\label{eq:pW}
p(T,\lambda) = \frac{T}{b} \, W[b \, \phi(T) \, \lambda]~,
}
in terms of the Lambert W-function~\cite{LambertW}
defined by the equation
\eq{\label{W}
 z ~ = ~W(z) \, \exp[ W(z)] 
 }
for any complex number $z$. Therefore, the representation (\ref{eq:pW})
provides the analytic continuation of the EV model pressure function into the complex fugacity plane.
$W(z)$ is, in general, a multi-valued function.
On the principal branch, $W(z)$ is real for real values of the argument $z$, and $W(z) \sim z$ for small values
of $z$. The principal branch therefore determines the physical behavior of the EV pressure at real, positive values of the fugacity $\lambda$.
In the following we consider the principal branch of $W(z)$ only.
This principal branch has the following Taylor series representation:
\eq{\label{eq:WTaylor}
W(z) = \sum_{k=1}^{\infty} \frac{(-k)^{k-1}}{k!} \, z^k~,
}
which follows from the Lagrange inversion theorem, applied to Eq.~(\ref{W}).
The fugacity expansion around $\lambda = 0$  of the pressure in the EV model therefore reads
\eq{\label{eq:pWTaylor}
p(T,\lambda) = \sum_{k=1}^{\infty} \frac{(-k)^{k-1}}{k!} \, T \, b^{k-1} \, [\phi(T)]^k \, \lambda^k~.
}

The pressure of the ideal Boltzmann gas~($b = 0$),
\eq{
p^{\rm id} (T,\lambda)~ = ~T \, \phi(T) \, \lambda~=~ nT~,
}
corresponds to the first term of the fugacity expansion~\eqref{eq:pWTaylor}.

It is instructive to consider the ratio $R$ of the EV pressure to the ideal gas pressure:
\eq{\label{eq:K}
R \equiv \frac{p(T,\lambda)}{p^{\rm id}(T,\lambda)} = \frac{W[b \, \phi(T) \, \lambda]}{b \, \phi(T) \, \lambda}~.
}

\begin{figure*}[t]
  \centering
  \includegraphics[width=.49\textwidth]{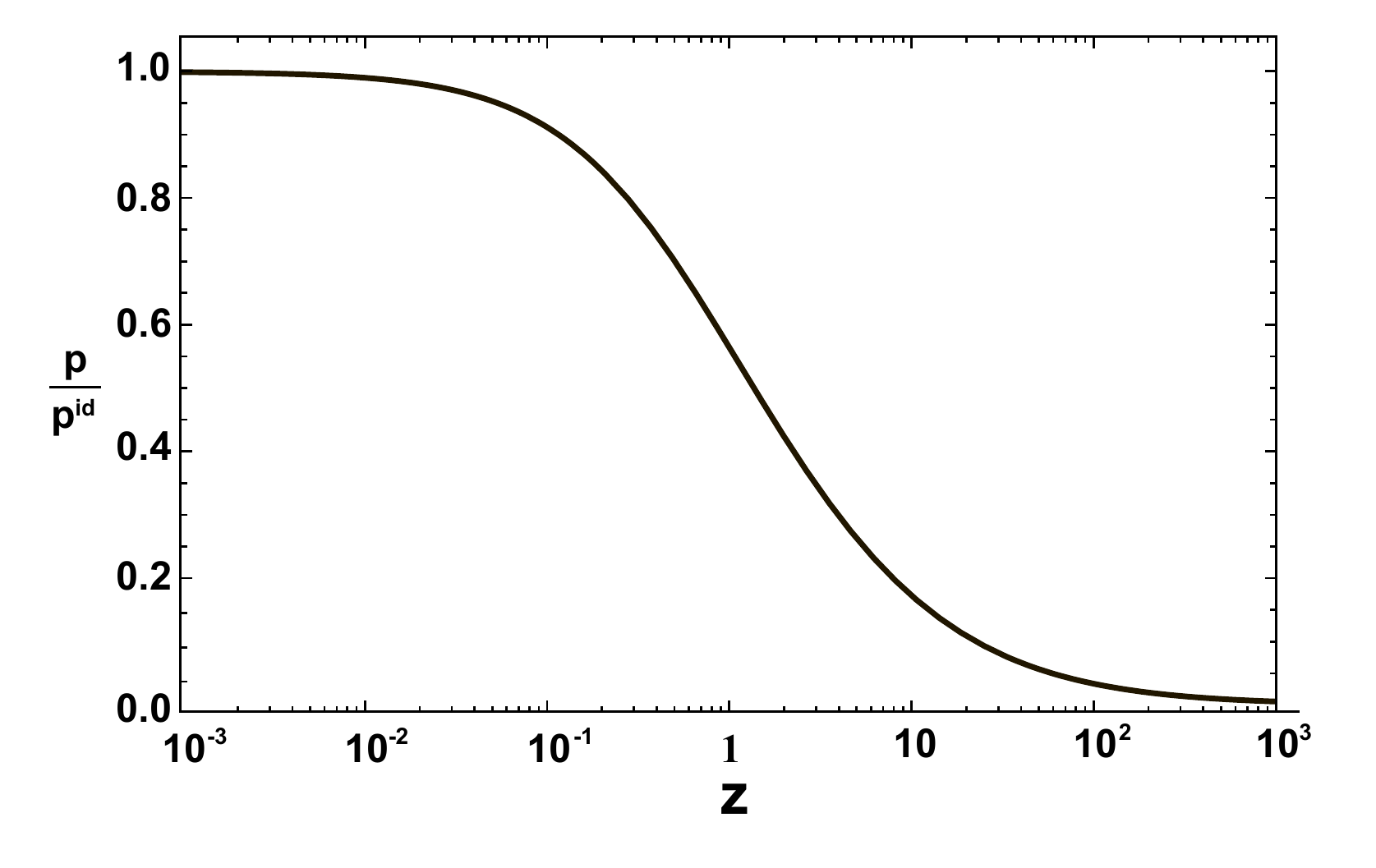}
  \includegraphics[width=.49\textwidth]{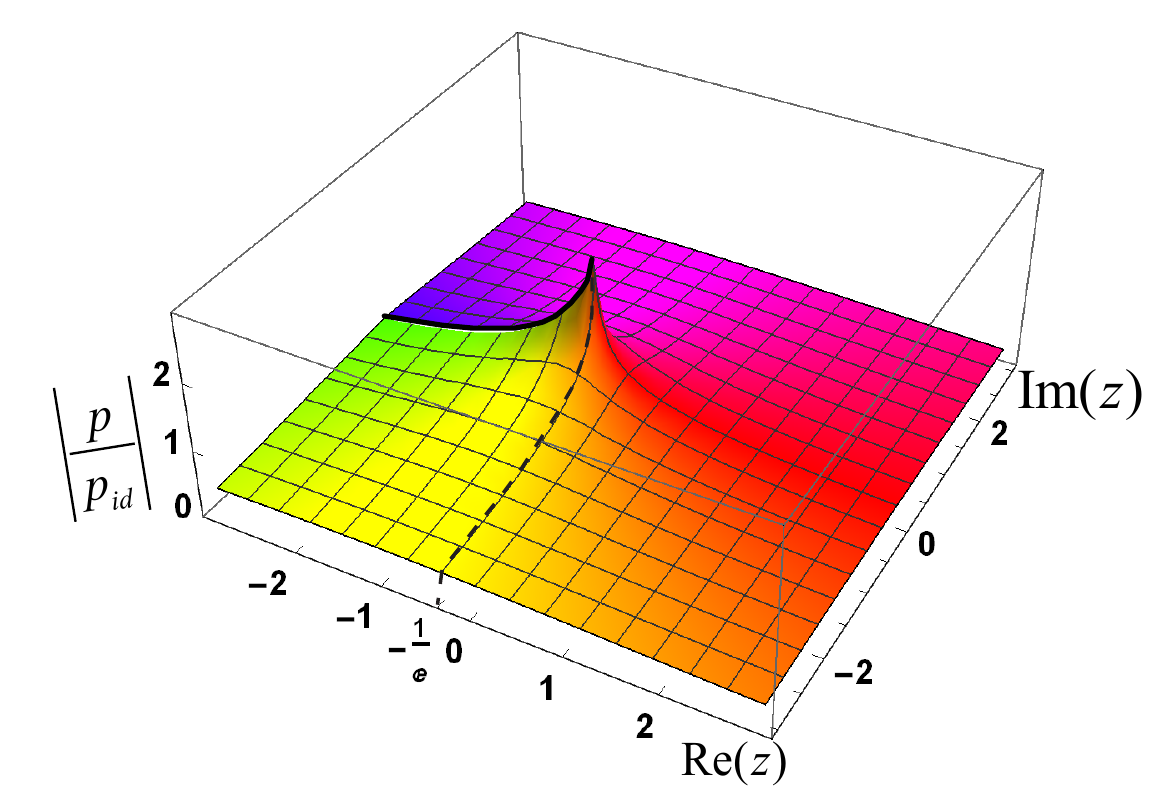}
  \caption{
   \emph{Left panel}: The dependence of the excluded-volume to ideal gas pressure ratio $R(z) \equiv p / p_{\rm id}$~[Eq.~\eqref{eq:K}] on the dimensionless fugacity $z$ shown on the logarithmic scale for real positive values of $z$. 
   \emph{Right panel}: The contour plot of $|R(z)|$ in the complex $z$ plane.
   The branch cut from $z = -\infty$ to $z = -e^{-1}$ is shown by the solid line while the dashed line corresponds to $\text{Re}~z = -e^{-1}$.
   The coloring denotes the phase angle of $R(z)$.
  }
  \label{fig:K}
\end{figure*}

This ratio quantifies the deviations from the ideal gas case. 
It depends on the dimensionless variable $z = b \, \phi(T) \, \lambda$ only, i.e. $R \equiv R(z) = W(z)/z$.
The $R(z)$ dependence is shown for real values of $z>0$ in Fig.~\ref{fig:K} (left panel).
The EV effects are moderate~(within 10\%) for $z \lesssim 10^{-1}$.
At $z \gtrsim 1$ the EV effects are quite strong, and the ideal gas picture simply breaks down.
These observations are useful as a rule of thumb, to estimate the importance of the EV corrections in various settings, e.g. as in the application of the EV model to the HRG phenomenology.

The contour plot of $|R(z)|$ in the complex $z$-plane is shown in Fig.~\ref{fig:K} (right panel):
$R(z)$ exhibits a branch point at $z = z_{\rm br} \equiv -e^{-1}$, with a branch cut along the interval $(-\infty, -e^{-1})$,
which follows from the analytic properties of the Lambert W-function. This branch cut is depicted by the black line.
Note that the Lee-Yang zeroes of the EV model are distributed along this branch cut~\cite{hauge1963yang}.
$|R(z)|$ is a continuous function of the complex-valued $z$, but the imaginary part of $R(z)$ flips its sign when crossing the branch cut.
$|R(z)| \to e$ as $z \to z_{\rm br}$.

\section{Hadron resonance gas with repulsive baryonic interactions}
\label{sec:evhrg}

\subsection{Excluded volume HRG model}

The EV approach is often applied to include repulsive interactions between hadrons in the HRG model.
% of hadronic equation of state.
The HRG model with EV interactions between (anti)baryons~(the EV-HRG model) was developed in Refs.~\cite{Vovchenko:2016rkn,Satarov:2016peb,Vovchenko:2017xad}.
This model treats the interactions between pairs of baryons and between pairs of anti-baryons, but not between any other pairs of hadrons, by excluded volume (EV) correction \'a la van der Waals.
These interactions are quantified by vdW-type eigenvolume parameter $b$.
The pressure in the EV-HRG model reads
\eq{\label{eq:pevhrg}
p(T,\lambda_B) = p_M(T) + p_B(T,\lambda_B) + p_{\bar{B}}(T,\lambda_B),
}
where $\lambda_B=\exp(\mu_B/T)$ and $\mu_B$
is the baryonic chemical potential.
Here
\eq{\label{eq:pev}
& p_{_{M}} (T) = T \, \phi_M(T)~, \\
& p_{_{B}} (T, \lambda_B) = T \, \phi_B(T) \, \lambda_B \,
\exp\left( \frac{- b \, p_{_{B}}}{T} \right)~,  \\
& p_{_{\bar{B}}} (T, \lambda_B) = T \, \phi_B(T) \, \lambda_B^{-1} \,
\exp\left( \frac{- b \, p_{_{\bar{B}}}}{T} \right)~, \\
%}
% Here
% \eq{
& \phi_{M(B)} (T)  = \sum_{i \in M(B)}  \int d m \, \rho_i(m) \, \frac{d_i  m^2  T}{2\pi^2}  K_2\left( m \over T \right), \label{phi}
}
where $\rho_i(m)$ in Eq.~(\ref{phi})
takes into account the finite widths of the resonances while the sum runs over all mesonic~(M) or baryonic~(B) species.

The previous section has shown that the explicit form of the pressure in the EV-HRG model is given in terms of the Lambert W-function:
\eq{\label{eq:pEVW}
& p(T,\lambda_B)  = T \, \phi_M(T) \nonumber \\
& + \frac{T}{b} \, \left\{ W[b \, \phi_B(T) \, \lambda_B] + W[b \, \phi_B(T) \, \lambda_B^{-1}] \right\}~.
}

\subsection{Taylor expansion properties}
\begin{figure}[t]
  \centering
  \includegraphics[width=.44\textwidth]{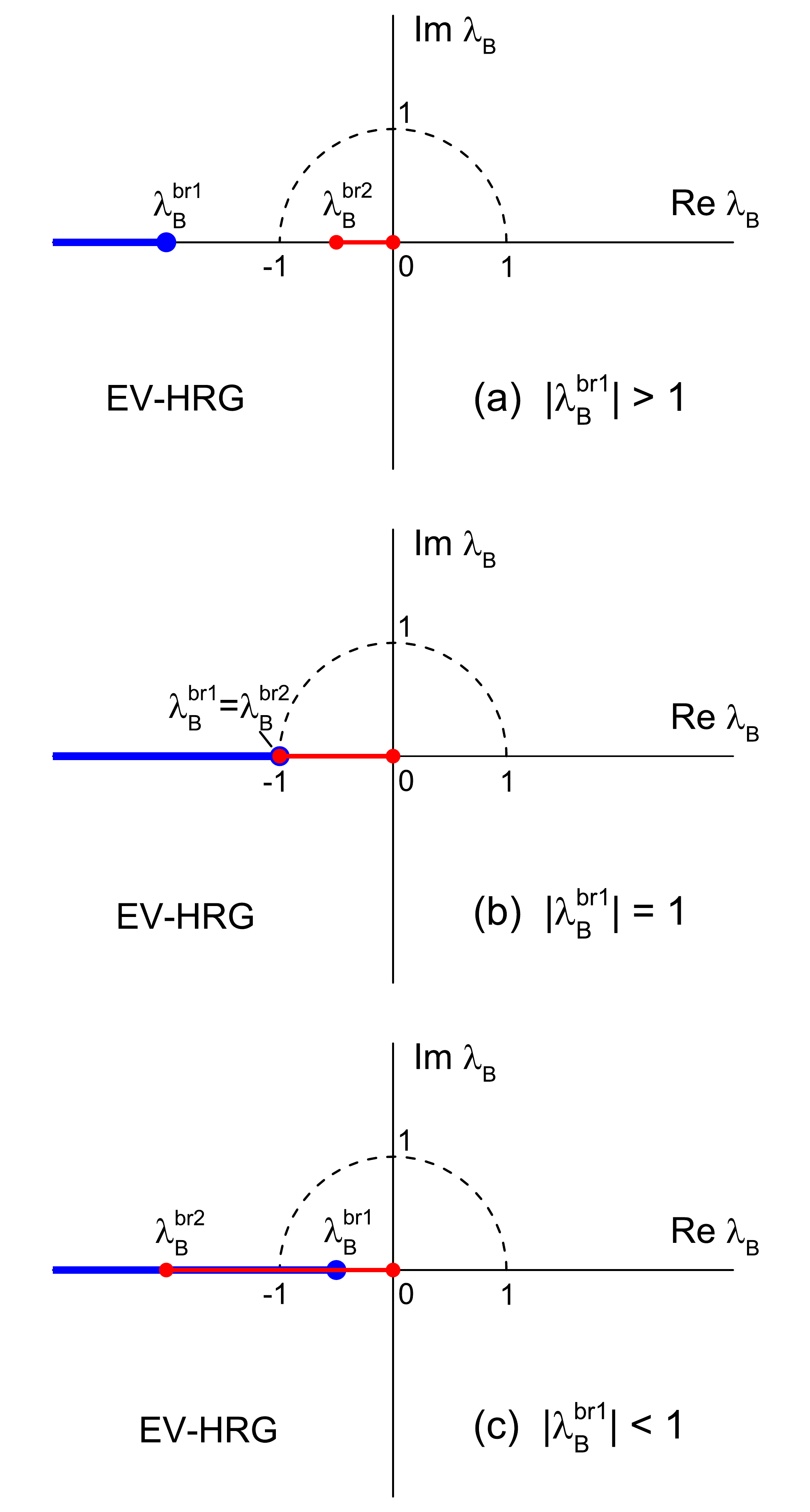}
  \caption{
   The analytic structure of the EV-HRG model pressure function is depicted in the complex fugacity plane for (a) $|\lambda_B^{\rm br1}| > 1$, (b) $|\lambda_B^{\rm br1}| = 1$, and (c) $|\lambda_B^{\rm br1}| < 1$.
   The blue and red lines with the points depict the branch cuts, the blue one corresponds to the branch cut in the 2nd term of Eq.~\eqref{eq:pEVW} and the red one to the branch cut in the 3rd term of Eq.~\eqref{eq:pEVW}.
   The dashed curves correspond to purely imaginary values of the baryochemical potential in the range $0 < \text{Im}~[\mu_B/T] < \pi$, the integration contour in Eq.~\eqref{eq:FourierDef}.
  }
  \label{fig:cuts}
\end{figure}

The branch points of the pressure
function (\ref{eq:pEVW}),
\eq{\label{eq:EVHRGbr}
\lambda_B^{\rm br 1,2} (T) = [-b \, \phi_B(T) \, e]^{\mp 1}~,
}
are located exclusively at the negative real axis.
Here $\lambda_B^{\rm br 1}$ corresponds to the branch point associated with the subsystem of baryons~[the second term in Eq.~\eqref{eq:pEVW}], while $\lambda_B^{\rm br 2}$ corresponds to the subsystem of antibaryons~[the third term in Eq.~\eqref{eq:pEVW}].
The two singularities with positions related as $\lambda_B^{\rm br 1}=\frac{1}{\lambda_B^{\rm br 2}}$ emerge due to the presence of both, baryons and antibaryons, which leads to two different branch cuts, both located at the negative real fugacity axis.
These branch cuts are depicted in Fig.~\ref{fig:cuts} for three different cases: 

\begin{enumerate}[label=(\alph*)]
\item $|\lambda_B^{\rm br1}| > 1$: the branch cuts do not overlap;

\item $|\lambda_B^{\rm br1}| = 1$: the two branch points coincide, $\lambda_B^{\rm br 1} = \lambda_B^{\rm br 2} = -1$;

\item $|\lambda_B^{\rm br1}| < 1$: the branch cuts have a non-zero overlap.
\end{enumerate}

The locations of the distinct branch points are given (for $k \in \mathcal{Z}$) in terms of the baryochemical potential:
\eq{\label{eq:EVHRGbrmu}
\frac{\mu_B^{\rm br} (T)}{T} = \pm \{\ln[b \, \phi_B(T)] + 1\} \pm i \, \pi \, (2k+1)~.
}

The pressure function (\ref{eq:pEVW}) can now be written as a Taylor series expansion around $\mu_B/T = 0$:
\eq{\label{eq:TaylorExp}
p(T,\mu_B) = p(T,\mu_B = 0) + \sum_{k=1}^{\infty} \, \frac{\chi_{2k} (T)}{(2k)!} \, \left( \frac{\mu_B}{T}\right)^{2k}~.
}
Here the coefficients of the expansion are the baryon number susceptibilities $\chi_{2k} (T) = \partial^{2k} (p/T^4) / \partial (\mu_B/T)^{2k}|_{\mu_B = 0}$, evaluated at $\mu_B = 0$.

The presentation (\ref{eq:TaylorExp}) is quite general and is applied here for the QCD equation of state.
The leading susceptibilities have been computed in lattice QCD simulations.
Current data are available for susceptibilities up to $\chi_8^B$~\cite{Bazavov:2017dus,Borsanyi:2018grb}.
Due to the CP-symmetry of QCD, all odd order susceptibilities vanish at $\mu_B = 0$.

The radius of convergence of the Taylor expansion~\eqref{eq:TaylorExp} is determined by the singularity
of the pressure function in the complex $\mu_B/T$ plane, which is located in the closest to the expansion point, $\mu_B/T = 0$.

Thermodynamic singularities are often associated with phase transitions.
For example, the critical endpoint of a first-order phase transition manifests itself as a singularity at real finite $\mu_B^{\rm crit}$,
which limits the convergence of the Taylor expansion around $\mu_B = 0$~\cite{GNS}.
This fact has been used in various  attempts to constrain the location of the critical point of QCD by numerical evaluation of a few leading  coefficients with lattice QCD~\cite{Allton:2002zi,Gavai:2004sd,Allton:2005gk} or in effective models~\cite{Stephanov:2006dn,Karsch:2010hm,Skokov:2010uc}.

The EV-HRG model (\ref{eq:EVHRGbrmu}) exhibits no physical phase transition.
Thus, it does not have singularities at real values of the baryochemical potential.
Nevertheless, the model does contain branch point singularities in the complex plane, their locations are given by Eq.~\eqref{eq:EVHRGbrmu}.
The closest branch points to $\mu_B / T = 0$ result by setting $k=0$ in Eq.~\eqref{eq:EVHRGbrmu}:
\eq{\label{mubr}
\frac{\mu_B^{\rm br} (T)}{T} = \pm\{1 + \ln[b \, \phi_B(T)]\} \pm i \, \pi~.
}
These two branch points are symmetric with respect to $\mu_B/T=0$, which reflects the symmetry between baryons and antibaryons.
The radius of convergence, $r_{\mu/T}$, of the Taylor expansion in the EV-HRG model is given by the distance of these
symmetric branch points to $\mu_B/T = 0$:
\eq{\label{rmu}
r_{\mu/T} = \sqrt{ \{1 + \ln[b \, \phi_B(T)] \}^2 + \pi^2 }~.
}

The Taylor expansion~\eqref{eq:TaylorExp} does converge only in the region $|\mu_B| / T < r_{\mu/T}$.
For illustration, the behavior of the Taylor expansion (\ref{eq:TaylorExp}) is studied in the EV-HRG model, where as an example $T = 155$~MeV and the model parameters from Ref.~\cite{Vovchenko:2017xad} are used:
\eq{\label{eq:Tparam}
b = 1~\text{fm}^3, \quad b \, \phi_B(T=155~\text{MeV}) \simeq 0.026~.
}
This yields the branch points (\ref{mubr}):
%of the partition function are located at
\eq{
\frac{\mu_B^{\rm br}}{T} \simeq \pm 2.634 \pm i \, \pi~,
}
while the radius of convergence (\ref{rmu}) becomes equal
\eq{
r_{\mu/T} ~ \simeq 4.1,
}
i.e. $r_{\mu} \simeq 635$~MeV.

\begin{figure}[t]
  \centering
  \includegraphics[width=.49\textwidth]{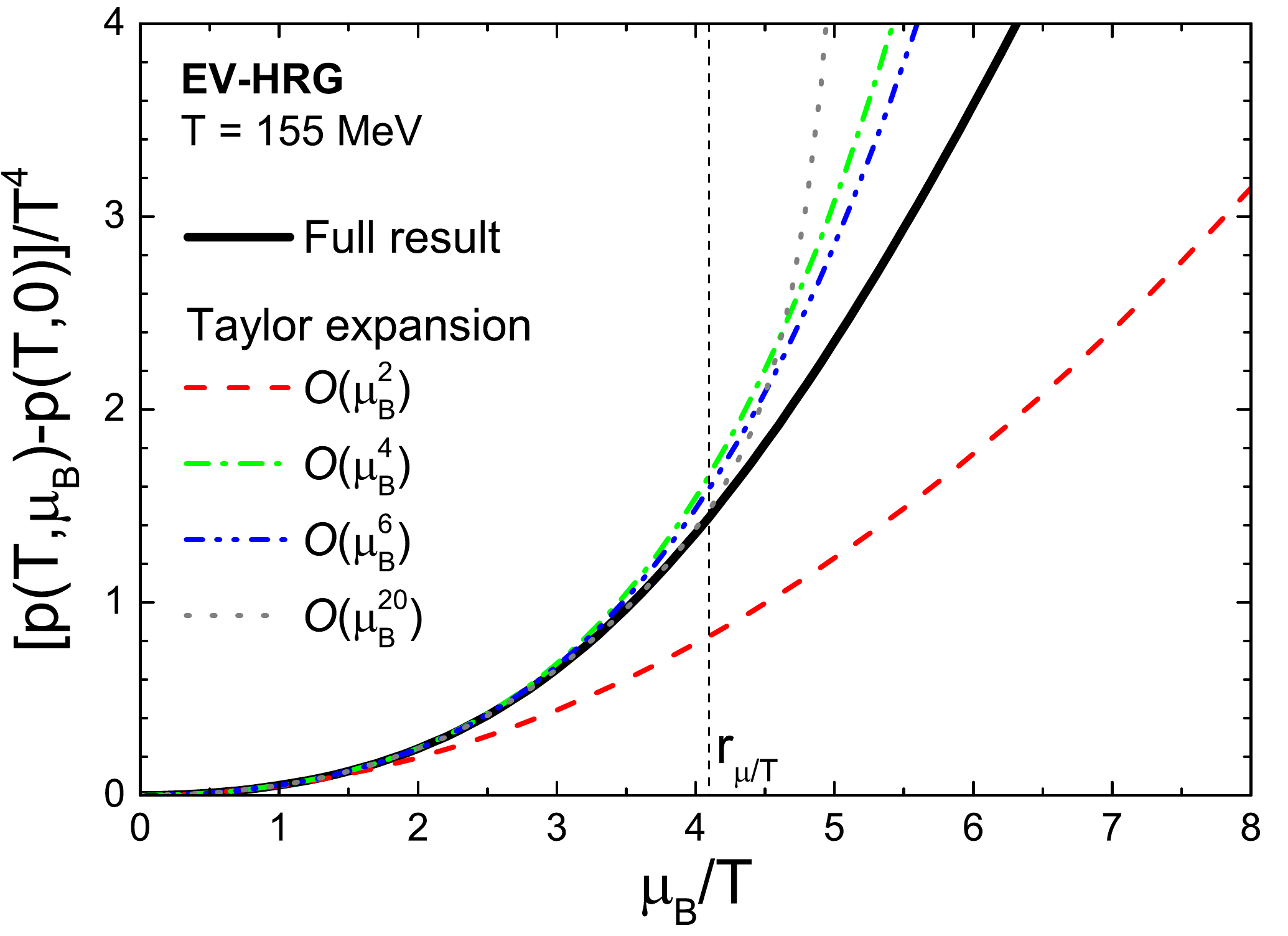}
  \caption{The dependence of the  subtracted scaled pressure $[p(T,\mu_B)-p(T,0)]/T^4$ on $\mu_B/T$, as calculated within the EV-HRG model at $T = 155$~MeV using the analytic solution~[Eq.~\eqref{eq:pevhrg}]~(solid black line) and the Taylor expansion truncated at $\chi_2^B$~(dashed red line), $\chi_4^B$~(dot-dashed green line), $\chi_6^B$~(double-dot-dashed blue line), and $\chi_{20}^B$~(dotted grey line).
  The vertical dashed line corresponds to the value of the convergence radius $r_{\mu/T} \simeq 4.1$.}
  \label{fig:TaylorTest}
\end{figure}

Figure~\ref{fig:TaylorTest} depicts the $\mu_B/T$ dependence of the subtracted
scaled pressure $[p(T,\mu_B)-p(T,0)]/T^4$, evaluated within the EV-HRG model at $T = 155$~MeV. 
The full analytic solution~[Eq.~\eqref{eq:pevhrg}]~(solid black line) is shown, as well as the Taylor expansion~\eqref{eq:TaylorExp},
truncated at $\chi_2^B$~(dashed red line), $\chi_4^B$~(dot-dashed green line), $\chi_6^B$~(double-dot-dashed blue line), and $\chi_{20}^B$~(dotted grey line).
The full analytic result is described fairly well by the Taylor expansion, if it is truncated at the $\mathcal{O}(\mu_B^4)$ order or higher, for $\mu_B/T \lesssim r_{\mu/T} \simeq 4.1$.
However, the behavior of the pressure function cannot be reliably described beyond the convergence radius by a truncated Taylor expansion, no matter how high is its order.
Moreover, the agreement of the partial sums in Eq.~(\ref{eq:TaylorExp}) with the exact result becomes, with an increasing number of their terms, better at  $\mu_B/T < r_{\mu/T}$, but worse at $\mu_B/T > r_{\mu/T}$ outside the convergence radius, as can be seen in Fig.~\ref{fig:TaylorTest}.
Note that the divergence of the Taylor expansion at large real $\mu_B/T > r_{\mu/T}$ does not at all indicate an emergence of physical effects. That observation does simply reflect the existence of complex chemical potential singularities, which limit the convergence range of a Taylor series.

The present results illustrate that the application of the Taylor expansion method in lattice QCD shall respect these findings and must be done carefully.
The convergence ranges of the Taylor expansion method are often restricted just by pressure function singularities, which are not at all related to physical phase transitions.

\subsection{Fourier coefficients}

The QCD net baryon density $n_B$
can be written as a series in hyperbolic sines,
\eq{\label{eq:rhoBFourierDef}
\frac{n_B(T,\mu_B)}{T^3} = \sum_{k=1}^{\infty} \, b_k(T) \, \sinh\left( \frac{k \mu_B}{T} \right)~.
}
This general representation is a consequence of the CP- and Roberge-Weiss~\cite{Roberge:1986mm} symmetries of QCD.
For purely imaginary chemical potentials $\mu_B$, this expansion becomes trigonometric Fourier series. Here the coefficients $b_k(T)$ are the Fourier coefficients,
which can be evaluated in the standard way:
\eq{\label{eq:FourierDef}
b_k(T) = \frac{2}{\pi} \int_0^{\pi} \text{Im} \left[ \frac{n_B(T, i \theta_B \, T)}{T^3} \right] \, \sin(k \, \theta_B) \, d \theta_B~.
}
These Fourier coefficients have attracted considerable attention  recently~\cite{Vovchenko:2017gkg,Almasi:2018lok,Bzdak:2018zdg,Almasi:2019bvl}, in particular in the context of lattice QCD simulations at imaginary $\mu_B$~\cite{Lombardo:2006yc,Bornyakov:2016wld,Vovchenko:2017xad}.

The four leading coefficients were analyzed in Ref.~\cite{Vovchenko:2017gkg} within the EV-HRG model, in the context of lattice data.
The analytic expression~\eqref{eq:pEVW} determines the exact expressions for $b_k$ to arbitrary order in the EV-HRG model.
The Taylor expansion of $W(z)$~[Eq.~\eqref{eq:WTaylor}] yields
\eq{\label{eq:pEVHRGser}
& p(T,\lambda_B)  = p_M(T) + \sum_{k=1}^{\infty} \, \frac{(-k)^{k-1} \, T \, b^{k-1} \, [\phi_B(T)]^k}{k!} \,  \lambda_B^n \nonumber \\
&  + \sum_{k=1}^{\infty} \, \frac{(-k)^{k-1} \, T \, b^{k-1} \, [\phi_B(T)]^k}{k!} \,  \lambda_B^{-k}
%\nonumber \\
  \nonumber \\
&  = p_M(T) + \sum_{k=1}^{\infty} \, \frac{(-k)^{k-1} \, 2 \, T \, b^{k-1} \, [\phi_B(T)]^k}{k!}  \, \cosh\left( \frac{k \mu_B}{T} \right)~.
}

The scaled net baryon density $n_B/T^3 = \partial(p/T^4) / \partial (\mu_B/T)$ reads
\eq{\label{eq:rhoEVHRGser}
\frac{n_B}{T^3} & = \sum_{k=1}^{\infty} \, \frac{(-k)^{k-1}}{k!} \, 2 \, k  \, \frac{b^{k-1}}{T^3} \, [\phi_B(T)]^k \, \sinh\left( \frac{k \mu_B}{T} \right)~, 
}
with the Fourier coefficients
\eq{\label{eq:bnEV}
b_k^{\rm ev}(T) = (-1)^{k-1} \, \frac{2 \, k^k}{k!} \, \frac{\phi_B(T)}{T^3}  [b \, \phi_B(T)]^{k-1} \,~.
}

The four leading Fourier coefficients read
\eq{
b_1^{\rm ev}(T) & = 2 \, \frac{\phi_B(T)}{T^3}\, , \\
b_2^{\rm ev}(T) & = -4 \, b \, T^3 \,  \left[ \frac{\phi_B(T)}{T^3} \right]^2 \, , \\
b_3^{\rm ev}(T) & = 9\, (b \, T^3)^2 \,  \left[ \frac{\phi_B(T)}{T^3} \right]^3 \, , \\
b_4^{\rm ev}(T) & = -\frac{64}{3} \, (b \, T^3)^3 \,  \left[ \frac{\phi_B(T)}{T^3} \right]^4 \, .
}
They agree with those obtained in Ref.~\cite{Vovchenko:2017xad}.
The closed-form expression~\eqref{eq:bnEV} suggests that the alternating sign structure of the Fourier coefficients in the EV-HRG model persists to asymptotically large $k$.
The Stirling approximation $k! \approx \sqrt{2\pi k} \, (k/e)^k$ yields the following large $k$  asymptotics:
\eq{\label{eq:bnEVasympt}
b_k^{\rm ev} \stackrel{k \to \infty}{\simeq} -\frac{\sqrt{2/\pi}}{b \, T^3} \, \frac{[-b \, \phi_B(T) \, e]^k}{k^{1/2}} \sim \frac{[\lambda_B^{\rm br 1}(T)]^{-k}}{k^{1/2}}~.
}

The Fourier coefficients are exponentially damped, at large $k$, as long as the following condition is fulfilled:
\eq{\label{eq:convergence}
b \, \phi_B(T) < e^{-1} \quad \Longleftrightarrow \quad |\lambda_B^{\rm br 1}| > 1.
}
The corresponding analytic structure of the thermodynamic potential in this case is depicted in Fig.~\ref{fig:cuts}(a).

In contrast, Eq.~\eqref{eq:bnEVasympt} implies an exponential growth of the coefficients at large $k$ for $|\lambda_B^{\rm br 1}| < 1$.
Such a behavior contradicts the Riemann-Lebesgue lemma~\cite{RLlemma}, which stipulates that Fourier coefficients of any function which is integrable on the imaginary $\mu_B/T$ interval  $[0,\pi]$ vanish for large $k$, $b_k \stackrel{k \to \infty}{\to} 0$.
This contradiction appears to be related to the divergence of the series in Eq.~\eqref{eq:pEVHRGser} for purely imaginary values of the baryochemical potential, $|\lambda_B| = 1$, used to evaluate $b_k$.
In fact the integration endpoint $\theta_B = \pi$ in Eq.~\eqref{eq:FourierDef} for $|\lambda_B^{\rm br 1}| < 1$ lies on the branch cuts of both $W$-functions which enter Eq.~\eqref{eq:pEVHRGser}~[see Fig.~\ref{fig:cuts}(c)].
Therefore, Eq.~\eqref{eq:bnEV} is expected to coincide with the Fourier coefficients evaluated through~\eqref{eq:FourierDef} only when the condition~\eqref{eq:convergence} is fulfilled simultaneously.

The Fourier coefficients can be evaluated numerically through~Eq.~\eqref{eq:FourierDef} to cross-check these results with Eq.~\eqref{eq:bnEV}.
Both results agree for $|\lambda_B^{\rm br 1}| \geq 1$ only, but they disagree for $|\lambda_B^{\rm br 1}| < 1$.
For the latter case, the numerically calculated $b_k$, Eq.~\eqref{eq:FourierDef}, show an asymptotic behavior $b_k \sim (-1)^{k-1}/{k}$.

\section{Comparison to other approaches}
\label{sec:othermodels}

The EV model is only one particular framework to treat repulsive interactions between particles.
A comparison with the other approaches is instructive as it permits to establish  the analytic properties of the generic features of all distinct repulsive interaction models presented here.

\subsection{The mean-field approach}

In the simplest version of a MF approach the interactions between particles are modeled through a common shift of the single-particle energies which is proportional to the number density by $U = K \, n $~\cite{Olive:1980dy}.
The relations $K>0$ and $K<0$ correspond  to repulsive and attractive interactions, respectively.
Such an approach has recently been used to model repulsive baryonic interactions in the HRG in the context of the lattice data on baryon number susceptibilities~\cite{Huovinen:2017ogf}.
Similar results were achieved by the EV-HRG model~\cite{Vovchenko:2017xad}.
In case of the Maxwell-Boltzmann statistics, the particle number density $n$ of a single-component
system in the GCE is given by the following transcendental equation:
\eq{\label{eq:MFrho}
n(T,\lambda) = \phi(T) \, \lambda \, \exp\left(-\frac{K \, n}{T}\right)~.
}
The similarity of Eq.~\eqref{eq:MFrho} to the transcendental equation for the pressure~\eqref{eq:evol} in the EV model is evident.
The solution of~\eqref{eq:MFrho} is given in terms of the Lambert W-function:
\eq{
n(T,\lambda) = \frac{T}{K} \, W\left[ \frac{K \phi(T) \lambda}{T} \right]~.
}

The analytic properties of the MF model are determined by the analytic properties of the Lambert W-function, in close analogy to the EV model.
The branch point of the MF-model thermodynamic potential is located at
\eq{\label{eq:MF}
\lambda^{\rm br} = -\frac{T}{K \, \phi(T) \, e}~.
}
This singularity is located on the negative real axis for $K>0$~(repulsive mean field) and on the positive real axis for $K<0$~(attractive mean field).
This result suggests that strong attractive interactions can lead to experimentally observable physical singularities.

The MF model can be used to model repulsive interactions between pairs of baryons and between pairs
of antibaryons in the same fashion as was done in Sec.~\ref{sec:evhrg} for the EV model~(see~\cite{Huovinen:2017ogf} for details).
The resulting net baryon density reads ($K>0$):
\eq{\label{rhomf}
n_B^{\rm mf} (T, \lambda_B) = \frac{T}{K} \, \left\{W\left[ \frac{K \phi_B(T) \lambda_B}{T} \right] -  (\lambda_B \to \lambda_B^{-1}) \right\}~.
}
Similar to Eq.~(\ref{eq:EVHRGbr}) for EV interactions,
the MF model (\ref{rhomf}) used here possesses two branch points
\eq{
\lambda_B^{\rm br 1,2}~=~\left[-\,\frac{K}{T}\,\phi_B(T)e\,\right]^{\mp 1}
}
located at the negative real axis. 
Here $\lambda_B^{\rm br 1}$ corresponds to baryons and $\lambda_B^{\rm br 2}$ to antibaryons, as in the EV-HRG model before.
The Fourier coefficients of the net baryon density can be evaluated in the MF model using the Taylor series representation~\eqref{eq:WTaylor} of $W$:
\eq{\label{eq:bnMF}
b_k^{\rm mf}(T) = (-1)^{k-1} \, \frac{2 \, k^{k-1}}{k! \, K \, T^2} \, \left[\frac{K \, \phi_B(T)}{T}\right]^k \,~.
}
The asymptotic behavior is the following:
\eq{\label{eq:bnMFasympt}
b_k^{\rm mf} \stackrel{k \to \infty}{\simeq} -\frac{\sqrt{2/\pi}}{K \, T^2} \, \frac{[-K \, \phi_B(T)
 \, e / T]^k}{k^{3/2}} \sim \frac{\left[\lambda_B^{\rm br 1}\right]^{-k}}{k^{3/2}}~.
}
This asymptotic behavior is similar to the EV model. 
However, the MF model has a different power-law factor, namely $k^{-3/2}$, instead of $k^{-1/2}$ which appears in the EV model.
As in the EV model, Eqs.~\eqref{eq:bnMF} and \eqref{eq:bnMFasympt} are valid here for $|\lambda_B^{\rm br 1}| > 1$.

\subsection{The cluster expansion model}

The cluster expansion model (CEM) for the equation of state of QCD-matter at finite baryon density has been introduced recently in Refs.~\cite{Vovchenko:2017gkg,Vovchenko:2018zgt}.
Repulsive baryonic interactions are taken into account as well as the Stefan-Boltzmann limit of massless quarks at high temperatures. 
This provides a state-of-the-art description of the available lattice data on Fourier coefficients and baryon number susceptibilities.
The CEM net baryon density reads
\eq{
\frac{n_B(T,\lambda_B)}{T^3} & = -\frac{2}{27\pi^2} \, \frac{\hat{b}_1^2}{\hat{b}_2} \left\{ 4 \pi^2 \, [\operatorname{Li}_1(x_+) - \operatorname{Li}_1(x_-)]\right. \nonumber \\
& \quad \left. + 3 \, [\operatorname{Li}_3(x_+) - \operatorname{Li}_3(x_-)] \right\}
}
Here $\hat{b}_{1,2} = \displaystyle \frac{b_{1,2}(T)}{b_{1,2}^{\rm SB}}$, $x_{\pm} = \displaystyle - \frac{\hat{b}_2}{\hat{b}_1} \, \lambda_B^{\pm 1}$, $\textrm{Li}_s(z) = \displaystyle \sum_{k=1}^{\infty} \frac{z^k}{k^s}$ is the polylogarithm, and the
\eq{
b_k^{\rm {_{SB}}} = \frac{(-1)^{k-1}}{k} \, \frac{4 \, [3 + 4 \, (\pi k)^2]}{27 \, (\pi k)^2},
}
are the Fourier coefficients as evaluated in the Stefan-Boltzmann limit of massless quarks.

The analytic properties of the CEM are determined by the analytic properties of the polylogarithm. The branch points of the thermodynamic potential are located at
\eq{
\lambda_B^{\rm br 1,2} = \left[-\frac{\hat{b}_1}{\hat{b}_2}\right]^{\pm 1}~.
}
The singularities are located on the negative real axis, if $\hat{b}_1 / \hat{b}_2 > 0$. Lattice data suggests $\hat{b}_1 > 0$ and $\hat{b}_2 > 0$ for $T > 135$~MeV~\cite{Vovchenko:2017xad}~(lattice data are presently not yet available for $T < 135$~MeV).

The Fourier coefficients in the CEM read~(see~\cite{Vovchenko:2017gkg})
\eq{\label{eq:bnCEM}
b_k^{\rm cem} = b_k^{\rm SB} \, \frac{(\hat{b}_2)^{k-1}}{(\hat{b}_1)^{k-2}},
}
with the following asymptotic behavior:
\eq{\label{eq:bnCEMasympt}
b_k^{\rm cem} \stackrel{k \to \infty}{\simeq} \frac{16 \, \hat{b}_1}{27} \, \frac{[-\hat{b}_2/\hat{b}_1]^{k-1}}{k} \sim \frac{\left[\lambda_B^{\rm br 1}\right]^{-k}}{k}~.
}
This asymptotic behavior is similar to the EV and MF models discussed above, but has a power-law factor of $k^{-1}$, instead of the $k^{-1/2}$ factor shown for the EV model or the $k^{-3/2}$ factor in the MF model.

\subsection{The ideal Fermi gas}

The Fermi-Dirac and Bose-Einstein quantum statistical effects can be associated with ``effective'' repulsive~(fermions) or attractive~(bosons) interactions~\cite{GNS}.
We analyze the analytic properties of the thermodynamic potential of both the ideal Fermi gases of the baryons and of the antibaryons.
The GCE
expression for the net baryonic number density of a relativistic ideal Fermi gas of degeneracy $d$ and mass $m$ is presented as~\cite{GNS}
\eq{\label{eq:FDBE}
& n_{B}  = \frac{d}{2\pi^2} \, \int_0^{\infty} k^2 \, dk \, \Big\{\left[ \lambda_B^{-1} \, \exp\left(\frac{\sqrt{k^2+m^2}}{T}\right) + 1 \right]^{-1} 
\nonumber \\
& - \left(\lambda_B\rightarrow \lambda_B^{-1}\right) \Big\} \nonumber \\
& = \sum_{k = 1}^{\infty} \, (-1)^{k-1} \, \frac{d \, m^2 \, T}{2\pi^2 \, k} \, K_2(k \, m/T) \, \left(\lambda_B^k -\lambda_B^{-k} \right)~.
}
The series representation in the last line of Eq.~\eqref{eq:FDBE} is valid for $m>0$.

The net baryonic density of the ideal Fermi gas has two singularities at
\eq{\label{lamfer}
\lambda_B^{\rm br 1,2}~ =~ \left[-\,\exp(m/T)\right]^{\pm 1}~.
}
located on the real negative fugacity axis, where
the magnitude is determined by the mass of the particles, as follows from the integral representation in Eq.~(\ref{eq:FDBE}).

Note that the relativistic ideal Bose gas, as for example an ideal gas of positively and negatively charged pions,   
exhibits singularities in the fugacity $\lambda_Q$ connected to the conserved electric charge. 
These are located at the \emph{positive} axis
\eq{\label{lambos}
\lambda_Q^{\rm br 1,2}~ =~ \left[\,\exp(m_\pi/T)\right]^{\pm 1}~.
}
Therefore, the ideal Bose gas does exhibit real physical singularities, which are connected to the Bose-Einstein condensation.

The behavior of ideal Fermi~(Bose) gases is quite similar
to the corresponding MF model with a repulsive~(attractive) mean field~[see Eq.~\eqref{eq:MF}].
As stated, the singularities on the real axis evidently do correspond to the point of the onset of the Bose-Einstein condensation.
For fermions, this issue is more subtle, as the singularities found do correspond to complex values of the chemical potential.

The expansion~\eqref{eq:FDBE} allows to evaluate the Fourier coefficients of the net baryon density in an ideal gas of baryons and antibaryons:
\eq{\label{eq:bnQS}
b_k^{\rm qs} = (-\, 1)^{k-1} \, \frac{d \, m^2 \, T}{\pi^2 \, k} \, K_2(k \, m/T)~.
}
The asymptotic behavior of these Fourier coefficients is given by the following expression:
\eq{\label{eq:bnQSasympt}
b_k^{\rm qs} \stackrel{k \to \infty}{\simeq} \frac{(-\,1)^{k-1} \, 2\, d}{k^{3/2}} \, \left( \frac{mT}{2\pi}\right)^{3/2}
\exp \left(-\frac{k\,m}{T}\right)  \sim \frac{[\lambda_B^{\rm br 1}]^{-k}}{k^{3/2}}~.
}
This asymptotic behavior is exactly the same as the one found in the mean-field model~(\ref{eq:bnMFasympt}).

\subsection{Some remarks on the radius of convergence}

All models with repulsive interactions considered show very similar analytic structure of the thermodynamic potential. 
In all cases, the branch points are located at the negative real fugacity axis.
The radius of convergence of the Taylor expansion around $\mu_B/T = 0$ equals
\eq{
r_{\mu/T} = \sqrt{ \left(\ln |\lambda_{\rm br}^{1,2}|\right)^2 + \pi^2} 
}
in all these models.
Note that here $(\ln |\lambda_{\rm br}^1|)^2 = (\ln |\lambda_{\rm br}^2|)^2$, i.e. both branch points lie at the same distance from $\mu_B/T = 0$.
It is instructive to consider the behavior of $r_{\mu/T}$ in these various models.

\begin{figure}[t]
  \centering
  \includegraphics[width=.49\textwidth]{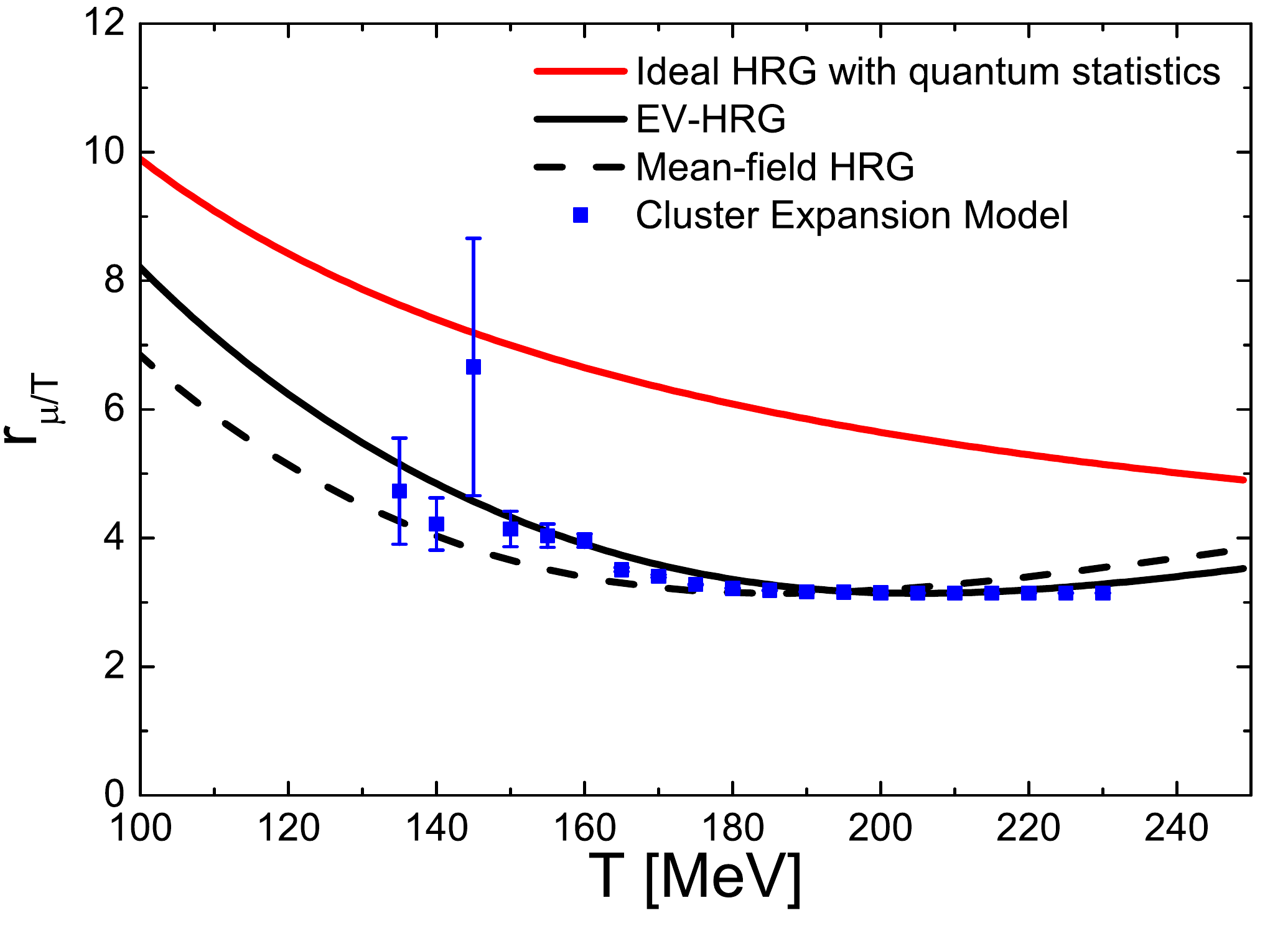}
  \caption{Temperature dependence of the radius of convergence of Taylor expansion around $\mu_B/T = 0$ evaluated for the ideal HRG model with quantum statistics~(solid red line), the EV-HRG model~(solid black line)~\cite{Vovchenko:2017xad}, the mean field HRG model~(dashed black line)~\cite{Huovinen:2017ogf}, and the cluster expansion model~(blue symbols with error bars)~\cite{Vovchenko:2017gkg}.
  }
  \label{fig:rmuTs}
\end{figure}

The radius of convergence in the ideal HRG model with quantum statistics is shown in Fig.~\ref{fig:rmuTs} by the red line as a function of temperature.
$r_{\mu/T}$ is defined there by the singularity in the Fermi-Dirac distribution function for nucleons and its value is determined by the vacuum mass of nucleons.

The $r_{\mu/T}$ values for EV-HRG and the MF-HRG models are shown in Fig.~\ref{fig:rmuTs} by the black solid and dashed lines, respectively.
Here we use $b = 1$~fm$^3$ for the EV-HRG model~\cite{Vovchenko:2017xad} and $K = 350$~MeV~fm$^3$ for the MF-HRG model~\cite{Huovinen:2017ogf}, reasonable parameter values suggested by comparisons to the lattice QCD data.
Both models predict similar values of $r_{\mu/T} \sim 3-5$ at $T>140$~MeV, reaching the minimum value of $r_{\mu/T}^{\rm min} = \pi$ at $T \simeq 190-200$~MeV. We do note that applicability of these hadron-based models might be questionable at high temperatures and our results there serve mainly for illustration purposes.
Similar values of $r_{\mu/T}^{\rm min}$ are predicted also by the CEM~(blue symbols in Fig.~\ref{fig:rmuTs})~\cite{Vovchenko:2017gkg}, where the lattice data for the two leading Fourier coefficients~\cite{Vovchenko:2017xad} were used as model input at each temperature value.
The radius of convergence in the CEM tends to $\pi$ at high temperatures, which may be associated with a Roberge-Weiss like transition~\cite{Roberge:1986mm}.
The results presented suggest that Taylor expansion is likely to be divergent at $\mu_B/T > 3-5$ and $T > 140$~MeV, regardless of existence of the hypothetical chiral critical point of QCD.

\subsection{Modeling the singular part of net baryon density with polylogarithms}

The asymptotic behavior of the Fourier coefficients in all examples considered has the form of an exponential decay times a power-law damping:
\eq{\label{eq:bkasymptgen}
b_k ~\stackrel{k \to \infty}{\sim}~\frac{\left[ \lambda_B^{\rm br 1}\right]^{-k}}{k^{\gamma}} \, \left[1 + \mathcal{O}\left(\frac{1}{k}\right) \right],
}
as long as $|\lambda_B^{\rm br1}| > 1$.
This asymptotic behavior is determined by a singularity of the net baryon density. 
The corresponding singular part of $n_B(T,\lambda_B)$ can then be approximated to the leading order:
\eq{
\frac{n_B^{\rm sing}(T,\lambda_B)}{T^3} \sim \sum_k \, \frac{(\lambda_B/\lambda_B^{\rm br1})^k + (\lambda_B/\lambda_B^{\rm br1})^{-k}}{k^\gamma}~,
}
as follows from the definition of the Fourier expansion for $n_B$~[see Eq.~\eqref{eq:rhoBFourierDef}].
Recalling the definition of the polylogarithm
\eq{
\operatorname{Li}_{\gamma} (z) = \sum_{k=1}^{\infty} \, \frac{z^k}{k^{\gamma}}~,
}
we arrive at
\eq{\label{eq:singpolylog}
\frac{n_B^{\rm sing}[T,\lambda_B]}{T^3}  \sim \left\{ \operatorname{Li}_{\gamma} [\lambda_B/\lambda_B^{\rm br1}] + \operatorname{Li}_{\gamma} \left[(\lambda_B/\lambda_B^{\rm br1})^{-1}\right] \right\}
}
as the leading order approximation of the singular part of the net baryon density in terms of the polylogarithm.
This approximation can be improved further on by considering the higher-order terms in the asymptotic expansion~\eqref{eq:bkasymptgen} for $b_k$, resulting in additional terms with polylogarithms of higher orders.

We considered an approximation of the Lambert W-function~(see the EV and MF models) in terms of the polylogarithms as described above as an example.
Namely, from an analysis of large $k$ terms in Eq.~\eqref{eq:WTaylor} it follows that
\eq{\label{eq:WvsLi}
W(z) \simeq -\frac{\operatorname{Li}_{3/2}(-z \, e)}{\sqrt{2\pi}} + \frac{\operatorname{Li}_{5/2}(-z \, e)}{12\sqrt{2\pi}} + \mathcal{O}(\operatorname{Li}_{7/2})~.
}
It is observed that a single polylogarithm $\operatorname{Li}_{3/2}$ can approximate $W(z)$ for $|z|<2$ to relative accuracy of better than 15\%, while the second-order approximation using two polylogarithms, $\operatorname{Li}_{3/2}$ and $\operatorname{Li}_{5/2}$, improves this accuracy to within 2\%.

The presented resummation of complex chemical potential plane singularities using polylogarithms is useful for phenomenological studies of thermodynamic singularities in QCD.

\section{Summary}
\label{sec:summary}

The analytic properties are studied within distinct approaches to treat repulsive interactions for the grand-canonical Maxwell-Boltzmann gas.
Main results are based on an observation that the EV model pressure can be expressed in terms of the Lambert W-function.
A single-component Maxwell-Boltzmann gas with an EV correction yields deviations from
the ideal gas behavior which depends universally on the dimensionless parameter $z = b \, \phi(T) \, \lambda$, where $b$
is the excluded-volume parameter, $\phi(T)$ is the ideal gas density at zero chemical potential, and $\lambda \equiv \exp(\mu/T)$ is the fugacity.

The analytic properties of the EV model are fully determined by the properties of the Lambert W-function. The pressure function of the EV model
has a regular behavior at all physical values of the fugacity/chemical potential,
but exhibits a branch cut singularity in the complex domain, namely at $\lambda^{\rm br} = [-b \, \phi(T) \, e]^{-1}$.
Therefore, the HRG
model with baryonic eigenvolumes has a finite radius of convergence of its Taylor
expansion around $\mu_B/T = 0$.  
This convergence radius is estimated to be $r_{\mu/T} \simeq 4.1$ for a crossover
transition temperature~($T\sim 155$~MeV), if a reasonable value is used for the baryonic excluded volume parameter, $b \simeq 1$~fm$^3$.

The Lambert W-function is used to determine the explicit form of the Fourier coefficients
of the  net baryon density~[Eq.~\eqref{eq:bnEV}], which shows an alternating sign behavior in all orders.
A number of other theories with repulsive interactions, such as the repulsive mean-field approach,
the cluster expansion model, and the ideal gas of fermions, show strong similarities of their analytic properties to the EV model.
In particular, the branch points of the pressure functions of all these approaches are all located on the
negative real fugacity axis. 
The asymptotic behavior of the Fourier coefficients does for all these models exhibit the form of an exponential decay 
times a power-law damping:
\eq{\label{eq:bksummary}
b_k ~\stackrel{k \to \infty}{\sim}~\frac{\left[ \lambda_B^{\rm br 1}\right]^{-k}}{k^{\gamma}} \quad \text{for} \quad |\lambda_B^{\rm br1}| > 1~.
}
The magnitude of the exponential suppression is in all cases 
universally determined by the location of the branch point of the pressure function which can be directly
associated with the repulsive interactions, whereas the power-law exponent $\gamma$ is specific to each model.
The alternating signs of the Fourier coefficients look the same in all considered examples and persist to asymptotically large $n$.
The universal asymptotic form~\eqref{eq:bksummary} allows to approximate the singular part of the net baryon density function in terms of polylogarithms, which is useful for phenomenological studies of thermodynamic singularities in QCD.

The present results are important in particular for the studies of the QCD phase structure,
this concerns both the lattice-based methods such as the Taylor expansion of the pressure in $\mu_B/T$, as well as the Fourier expansion of the net baryon density at imaginary chemical potential.
In fact, a pressure function singularity which can not be related to
a phase transition or a critical point does strongly restrict the convergence radius of the Taylor- and/or Fourier expansion methods.

The present work focuses on theories with  repulsive interactions only: Hence there is no possibility of a physical phase transition and/or a critical point.
It will be interesting to extend these studies within more elaborate phenomenological models of QCD
to determine the analytical structure of the pressure function for real values of the baryonic chemical potential,
e.g. for the hypothetical case where a phase transition occurs.

\section*{Acknowledgments}

The work of M.I.G. is supported by the Alexander von Humboldt Foundation and by the Program of Fundamental Research of the Department of Physics and Astronomy of the National Academy of Sciences of Ukraine
H.St. acknowledges the support through the Judah M. Eisenberg Laureatus Chair by Goethe University  and the Walter Greiner Gesellschaft, Frankfurt.

%\end{spacing}

\bibliography{EV-analyt}

\end{document}